# Investigation of $g_{f_0\omega\gamma}$ coupling constant in three point QCD sum rules and light cone sum rules

C Aydin[*] and A H Yilmaz[†]

*Physics Department, Karadeniz Technical University, 61080 Trabzon, Turkey*
[*] coskun@ktu.edu.tr
[†] hakany@ktu.edu.tr

**Abstract**

The coupling constant of $f_0 \to \omega\gamma$ decay is calculated using 3-point sum rule and light cone QCD sum rules. We investigate the results within the two-models which depend on $\theta$ angle.

## I. Introduction

The QCD sum rules [1] is one of the powerful methods for calculation of the different low energy characteristics of the hadron physics. In the context of QCD sum rules method, a wide class of problems of hadron physics, such as, spectrum, mass, weak form factors ([1-3], and references therein), magnetic moments of neutron and proton [4] etc., are successfully explained. Application of this method to polarization operators gives a determination of masses and couplings of low lying mesonic [1, 5] and baryonic [6] states. The QCD sum rule method has been utilized to analyze many hadronic properties, and it yields an effective framework to investigate the hadronic observables such as decay constants and form factors within the nonperturbative contributions proprotional to the quark and gluon condensates. [7]. The main idea of the method is to calculate the correlator with the help of operator product expansion (OPE) in the framework of QCD and then connect them with the phenomenological part. The interested physical quantities are determined by matching these two representations of the correlator.

Radiative transitions between pseudoscalar (P) and vector (V) mesons have been an important subject in low-energy hadron physics for more than three decades. These transitions have been regarded as phenomenological quark models, potential models, bag models, and effective Lagrangian methods [8,9]. Among the characteristics of the electromagnetic interaction processes $g_{VP\gamma}$ coupling constant plays one of the most important roles, since they determine the strength of the hadron interactions. In the quark models, V → P+γ decays (V = φ, ρ, ω) ; P = π, η, η′) are reduced by the quark magnetic moment with transition s = 1 → s = 0, where s is the total spin of the $q\bar{q}$ − system (in the corresponding meson). Actually these quantities can be calculated directly from QCD. Low-energy hadron interactions are governed by nonperturbative QCD so that it is very difficult to get the numerical values of the coupling constants from the first principles. For that reason a semiphenomenological method of QCD sum rules can be used, which nowadays is the standart tool for studying of various characteristics of hadron interactions. On the other hand, vector meson-pseudoscalar meson-photon VPγ–vertex also plays a role in photoproduction reactions of vector mesons on nucleons. It should be notable that in these decays (V→Pγ) the four-momentum of the pseudoscalar meson P is time-like, $P'^2 > 0$, whereas in the pseudoscalar exchange amplitude contributing to the photoproduction of vector mesons it is space-like $P'^2 < 0$. Therefore, it is of interest to study the effective coupling constant $g_{VP\gamma}$ from another point of view as well. In addition, the same model predicts amplitudes for energetically allowed $S \to V\gamma$ processes, for examples $f_0 \to \omega\gamma$, $f_0 \to \rho\gamma$, $a_0 \to \omega\gamma$ and $a_0 \to \rho\gamma$ etc. Black et all. [10] investigated the angular dependence on these decays.

The light scalar mesons have been the subject of continious interest in hadron spectroscopy. Although the structure of these light scalar mesons have not been unambiguously determined yet [11], the possibility may be suggested that these nine scalar mesons below 1 GeV may form a scalar SU(3) flavor nonet [12]. The nature of the meson $f_0(980)$ is particularly debated. One of the oldest suggestions, there is the proposal that quark confinement could be explained through the existence of a state with vacuum quantum numbers and mass close to the proton mass [13]. After having followed the quark model and considered the strong coupling to kaons, $f_0(980)$ could be interpreted as an $s\bar{s}$ state [14-17]. On the other hand, it does not explain the mass degeneracy between $f_0(980)$ and $a_0(980)$ interpreted as a $(u\bar{u} \pm d\bar{d})/\sqrt{2}$ state. A four quark $qq\bar{q}\bar{q}$ state definition has also been offered [18]. In this case, $f_0(980)$ could either be nucleon-like [19], i.e., a bound state of quarks with symbolic quark structure $f_0 = s\bar{s}(u\bar{u} + d\bar{d})/\sqrt{2}$, the $a_0(980)$ being $a_0 = s\bar{s}(u\bar{u} - d\bar{d})/\sqrt{2}$, or deuteron-like, i.e., a bound state of hadrons. The identification of the $f_0$ and of the other lightest scalar mesons with the Higgs nonet of a hidden $U(3)$ symmetry has also been proposed [20]. On the other hand, they are relevant hadronic degrees of freedom, and therefore the role they play in hadronic processes should also be studied besides the questions of theire nature. In this work, we calculated the coupling constant $g_{f_0\omega\gamma}$ by applying 3-point QCD sum rule and light-cone sum rule as well, which provide an efficient and model-independent method to study many hadronic observables, such as decay constants and form factors in terms of non-perturbative contributions proportional to the quark and gluon condensates [7].

## II. Calculation

### a) In the three-point QCD sum rules:

According to the general strategy of QCD sum rules method, the coupling constants can be calculated by equating the representations of a suitable correlator calculated in terms of hadronic and quark-gluon degrees of freedom. In order to do this we consider the following correlation function by using the appropriately chosen currents

$$\Pi_{\mu\nu}(p,p') = \int d^4x \, d^4y \, e^{ip'.y} e^{-ip.x} <0|T\{J_\mu^\gamma(0) J_{f_0}(x) J^\omega(y)\}|0> \qquad (1)$$

We choose the interpolating current for the $\omega$ and $f_0$ mesons as $j_\mu^{\omega_0} = \frac{1}{2}(\bar{u}\gamma_\mu^a u^a + \bar{d}\gamma_\mu^a d^a)$, and $J_{f_0} = \frac{1}{\sqrt{2}}(\bar{u}^b u^b + \bar{d}^b d^b)\sin\theta + \cos\theta \, s\bar{s}$ [21] respectively. $\omega$-meson consist of $u$ and $d$-quarks, we then ignore $s$-quark contribution in this calculation. $J_\mu^\gamma = e_u(\bar{u}\gamma_\mu u) + e_d(\bar{d}\gamma_\mu d)$ is the electromagnetic current with $e_u$ and $e_d$ being the quark charges.

The theoretical part of the sum rule in terms of the quark-gluon degrees of freedom for the coupling constant $g_{f_0\omega\gamma}$ is calculated by considering the perturbative contribution and the power corrections from operators of different dimensions to the three-point correlation function $\Pi_{\mu\nu}$. For the perturbative contribution we study the lowest order bare-loop diagram. Moreover, the power corrections from the operators of different dimensions $<\bar{q}q>$, $<\bar{q}\sigma.Gq>$, and $<(\bar{q}q)^2>$ are considered in the work. Since it is estimated to be negligible for light quark systems, we did not consider the gluon condensate contribution proportional to $<G^2>$. We performe the calculations of the power corrections in the fixed point gauge [22]. We also work in the limit $m_q = 0$ and in this limit the perturbative bare-loop diagram does not make any contribution. In fact, by considering this limit only operators of dimensions d=3 and d=5 make contributions which are proportional to $<\bar{q}q>$ and $<\bar{q}\eta.Gq>$, respectively. The relevant Feynman diagrams for power corrections are given in Fig 1.

On the other hand, in order to calculate the phenomenological part of the sum rule in terms of hadronic degrees of freedom, a double dispersion relation satisfied by the vertex function $\Pi_{\mu\nu}$ is considered [1, 2, 5]:

$$\Pi_{\mu\nu}(p,p') = \frac{1}{\pi^2} \int ds_1 \int ds_2 \frac{\rho_{\mu\nu}(s_1,s_2)}{(p^2-s_1)(p'^2-s_2)} \qquad (2)$$

where we ignore possible substruction terms since they will not make any contributions after Borel transformation. For our purpose we choose the vector and pseudoscalar channels and saturating this dispersion relation by the lowest lying meson states in these channels the physical part of the sum rule is obtained as

$$\Pi_{\mu\nu}(p,p') = \frac{<0|J^{f_0}|f_0><f_0(p)|J_\mu^\gamma|\omega(p')><\omega|J_\nu^\omega|0>}{(p^2-m_{f_0}^2)(p'^2-m_\omega^2)} + ..., \qquad (3)$$

where the contributions from the higher states and the continuum are given by dots. The overlap amplitudes for vector and pseudscalar mesons are $<0|J_\mu^\omega|\omega> = \lambda_\omega \varepsilon_\mu^\omega$, where $\varepsilon_\mu^\omega$ is the polarization vector of the vector meson and $<f_0|J_s|0> = \lambda_{f_0}$, respectively. The matrix element of the electromagnetic current is given by

$$<f_0(p)|J_\mu^\gamma|\omega(p')> = -i\frac{e}{m_{\omega_0}} g_{f_0\omega\gamma} K(q^2)(p.q\varepsilon_\mu - \varepsilon.qp_\mu) \tag{4}$$

where $q = p - p'$ and $K(q^2)$ is a form factor with $K(0) = 1$. This matrix element defines the coupling constant $g_{f_0\omega\gamma}$ by means of the effective Lagrangian

$$\mathcal{L} = \frac{e}{m_\omega} g_{f_0\omega\gamma} \partial_\mu \omega_\nu (\partial_\nu A_\beta - \partial_\beta A_\nu) f_0 \tag{5}$$

describing the $f_0\omega\gamma$ – vertex [23].

After performing the double Borel transform with respect to the variables $Q^2 = -p^2$ and $Q'^2 = -p'^2$, and by considering the gauge-invariant structure $(-g_{\mu\nu}(pq) + p_\mu q_\nu)$, we obtain the sum rule for the coupling constant as

$$g_{f_0\omega\gamma} = \frac{m_\omega}{2\sqrt{2}\lambda_\omega\lambda_{f_0}} e^{m_{f_0}^2/M_1^2} e^{m_\omega^2/M_2^2} <\bar{u}u> \left(3 - \frac{3}{8}\frac{m_0^2}{M_2^2} - \frac{7}{8}\frac{m_0^2}{M_1^2}\right)\sin\theta \tag{6}$$

where we used the relation $<\bar{q}\sigma_{\mu\nu}.G_{\mu\nu}q> = m_0^2 <\bar{q}q>$.

**b) In the light-cone sum rules:**

In order to derive the light cone QCD sum rule for the coupling constants $g_{f_0\omega\gamma}$, we consider the following two point correlation function

$$T_\nu(p,p') = i\int d^4x e^{ip'.x} \langle 0|T\{j_\nu^\omega(x) j_{f_0}(0)\}|0\rangle_\gamma, \tag{7}$$

where $\gamma$ denotes the external electromagnetic field, and $j_\nu^\omega$ and $j_{a_0}$ and are the interpolating current for the $\omega$ meson and $f_0$, respectively.

We therefore sature the dispersion relation satisfied by the vertex function $T_\mu$ by these lowest lying meson states in the vector and the scalar channels, and in this way we obtain for the physical part at the phenomenological level the Eq. (1) can be expressed as

$$T_\nu(p,p') = \frac{\langle 0|j_\nu^\omega|\omega\rangle\langle\omega(p')|f_0(p)\rangle_\gamma\langle f_0|j_{f_0}|0\rangle}{(p'^2 - m_\omega^2)(p^2 - m_{f_0}^2)}. \tag{8}$$

In this calculation the full light quark propagator with both perturbative and nonperturbative contribution is used, and it is given as [24]

$$iS(x,0) = \langle 0|T\{\bar{q}(x)q(0)\}|0\rangle$$

$$= i\frac{\not{x}}{2\pi^2 x^4} - \frac{<\bar{q}q>}{12} - \frac{x^2}{192}m_0^2 <\bar{q}q> - ig_s \frac{1}{16\pi^2}\int_0^1 du\left\{\frac{\not{x}}{x^2}\sigma_{\mu\nu}G^{\mu\nu}(ux) - 4iu\frac{x_\mu}{x^2}G^{\mu\nu}(ux)\gamma_\nu\right\} + ...$$

(9)

where the terms proportional to light quark mass $m_u$ or $m_d$ are neglected. After a straightforward computation we have

$$T_\mu(p,q) = 2i\int d^4 x e^{ipx} A(x_\rho g_{\mu\tau} - x_\tau g_{\mu\rho})\langle\gamma(q)|\bar{q}(x)\sigma_{\tau\rho}q(0)|0\rangle \tag{10}$$

where $A = \frac{i}{2\pi^2 x^4}$, and higher twist corrections are neglected since they are known to make a small contribution [25]. In order to evaluate the two point correlation function further, we need the matrix elements $\langle\gamma(q)|\bar{q}(x)\sigma_{\tau\rho}q(0)|0\rangle$. This matrix element can be expanded in the light cone photon wave function [26, 27]

$$\langle\gamma(q)|\bar{q}\sigma_{\alpha\beta}q|0\rangle = ie_q <\bar{q}q> \int_0^1 du e^{iuqx}$$

$$\times\{(\varepsilon_\alpha q_\beta - \varepsilon_\beta q_\alpha)[\chi\varphi(u) + x^2[g_1(u) - g_2(u)]] + [q.x(\varepsilon_\alpha x_\beta - \varepsilon_\beta x_\alpha) + \varepsilon.x(x_\alpha q_\beta - x_\beta q_\alpha)]g_2(u)\}$$

(11)

Where $e_q$ is the corresponding quark charge, $\chi$ is the magnetic susceptibility, $\varphi(u)$ is leading twist two and $g_1(u)$ and $g_2(u)$ are the twist four photon wave functions. The main difference between the tradiational QCD sum rules and light cone QCD sum rule is the appearence of these wave function. Light cone QCD sum rules corresponds to summation of an infinite set of terms in the expansion of this matrix element on the tradiational sum rules. The price one pays for this is the appearance of a priori unknown photon wave functions. After evaluating the Fourier transform for the $M_1$ structure and then performing the double Borel transformation with respect to the variables $Q_1^2 = -p^2$ and $Q_2^2 = -p'^2$, we finally obtain the following sum rule for the coupling constant $g_{f_0\omega\gamma}$

$$g_{f_0\omega\gamma} = \frac{1}{3\sqrt{2}}\frac{m_\omega <\bar{u}u>}{\lambda_{f_0}\lambda_\omega}e^{m_\omega^2/M_1^2}e^{m_{f_0}^2/M_2^2}\{-M^2\chi\phi(u_0)E_0(s_0/M^2) + 4g_1(u_0)\}\sin\theta \tag{12}$$

where the function

$$E_0(s_0/M^2) = 1 - e^{-s_0/M^2} \tag{13}$$

is the factor used to subtract the continuum, $s_0$ being the continuum threshold, and

$$u_0 = \frac{M_2^2}{M_1^2 + M_2^2}, \quad M^2 = \frac{M_1^2 M_2^2}{M_1^2 + M_2^2} \quad (14)$$

with $M_1^2$ and $M_2^2$ are the Borel parameters in the $\omega$ and $f_0$ channels.

## III. Numerical Calculation

From the 3-point QCD sum rules view, in our calculations we use the numerical values $<\bar{u}u> = -0.014$ GeV$^3$, $m_{f_0} = 0.98$ GeV, $\lambda_{f_0} = 0.18 \pm 0.02$ GeV$^2$ [28], $m_\omega = 0.782$ GeV. We note that neglecting the electron mass the $e^+e^-$ decay width of $\omega$ meson is given as $\Gamma(\omega \to e^+e^-) = \frac{4\pi\alpha^2}{3}\left(\frac{\lambda_\omega}{3}\right)^2$. Then using the value from the experimental leptonic decay width $\Gamma(\omega \to e^+e^-) = 0.60 \pm 0.02$ of keV for $\omega$ [29], we obtain the value $\lambda_\omega = (0.108 \pm 0.002)$ GeV$^2$ for the overlap amplitude $\omega$ meson. In order to examine the dependence of $g_{f_0\omega\gamma}$ on the Borel masses $M_1^2$ and $M_2^2$, we choose $M_1^2 = M_2^2 = M^2$. Since the Borel mass $M^2$ is an auxiliary parameter and the physical quantitites should not depend on it, one must look for the region where $g_{f_0\omega\gamma}$ is practically independent of $M^2$. We determined that this condition is satisfied in the interval $1.0 \text{ GeV}^2 \leq M^2 \leq 1.4 \text{ GeV}^2$. The variation of the coupling constant $g_{f_0\omega\gamma}$ as a function of Borel parameter $M^2$ at different $\theta$ values are shown in figure 2. Examination of this figure shows that the sum rule is rather stable with these reasonable variations of $M^2$. In the 3-point QCD sum rules calculation, we then choose the middle value $M^2 = 1.2 \text{ GeV}^2$ for the Borel parameter in its interval of variation and obtain the coupling constant of $g_{f_0\omega\gamma}$ for various $\theta$ angles as between $g_{f_0\omega\gamma} = 0.68 \pm 0.02$ and $g_{f_0\omega\gamma} = 1.25 \pm 0.02$, where only the error arising from the numerical analysis of the sum rule is considered.

From the light-cone sum rules view, we use the numerical values mentioned above as well as for the magnetic susceptibility $\chi = -3.15 \text{ GeV}^{-2}$ [30]. Using the conformal invariance of QCD up to one loop order, the photon wave functions can be expanded in terms of Gegenbauer polynomials; each term corresponding to contributions from operators of various conformal spin. Due to conformal invariance of QCD up to one loop, each term in this expansion is renormalized separately and the form of these wave functions at a sufficiently high scale is well known. In [26,27] it is shown that even at small scales, the wave functions do not deviate considerably from their asymptotic form [31, 32] and hence we will use the asymptotic forms of the photon wave function given by:

$$\varphi(u_0) = 6u_0(1-u_0),$$

$$g_1(u_0) = -\frac{1}{8}(1-u_0)(1-3u_0). \tag{15}$$

Since $m_{f_0} \approx m_\omega$, we will set $M_1^2 = M_2^2 = 2M^2$ which sets $u_0 = 1/2$. Note that in this approximation, we only need the value of the wave functions at a single point; namely at $u_0 = 1/2$ and hence the functional forms of the photon wave functions are not relevant.

In Fig. 3, we showed the dependence of the coupling constant $g_{f_0\omega\gamma}$ on parameter $M^2$ at constant value of the continuum threshold as $s_0 = 2.0$ at different $\theta$ values. In this case, we have coupling constant as between $g_{f_0\omega\gamma} = 0.78 \pm 0.02$ and $g_{f_0\omega\gamma} = 1.30 \pm 0.02$. In Fig. 4, we also showed the dependence of the coupling constant $g_{f_0\omega\gamma}$ on parameter $M^2$ at some different values of the continuum threshold as $s_0 = 1.8, \ 2.0$ and $2.2\,\text{GeV}^2$ at $\theta = 30°$. Since the Borel masses $M_1^2$ and $M_2^2$ are the auxiliary parameters and the physical quantities should not depend on them, one must look for the region where $g_{f_0\omega\gamma}$ is practically independent of $M_1^2$ and $M_2^2$. We determined that this condition is satisfied in the interval $1.2\,\text{GeV}^2 \leq M_1^2 \leq 1.4\,\text{GeV}^2$. Choosing the middle value $M^2 = 1.3\,\text{GeV}^2$ for the Borel parameter in this interval of variation and we have the coupling constant of $g_{f_0\omega\gamma}$ for different $s_0$ values as between $g_{f_0\omega\gamma} = 0.69 \pm 0.02$ and $g_{f_0\omega\gamma} = 0.71 \pm 0.02$ in the calculation of light-cone sum rules. The variation of the coupling constant $g_{f_0\omega\gamma}$ as a function of different values $M^2$ and $\theta$ are given in Fig. 5. Examination of this figure points out that the sum rule is rather stable with these reasonable variations of $M^2$.

## IV. Conclusions

In this study we calculated coupling constant $g_{f_0\omega\gamma}$ in the two different ways in which we took account $u$- and $d$-quark contribution. If our results are not agreed with the experimental ones, in this case one pays attention to s-quark contribution. In order to compare and to find the relation between the two models, we investigated the coupling constant within both approaches. In spite of lacking experimental data on $g_{f_0\omega\gamma}$, we found estimated values for the coupling constant $g_{f_0\omega\gamma}$ in three-point QCD sum rules and light-cone sum rules. The results depend on mixing angle $\theta$ and $s_0$ parameter. When one is used reasonable data respect to analytical expressions, it is clear that one has beter agreement to experimental results. For the time being there is no experimental data on $f_0\omega\gamma-$ vertex, then our calculations behave only a theoretical suggestion. However we offer that $f_0\omega\gamma-$ vertex in the related energy scala is very important such as $\phi\ f_0\gamma$ decay.

**Acknowledgments**

We are grateful to Prof. T.M. Aliev for careful reading of the manuscript and helpful suggestions about its revision.

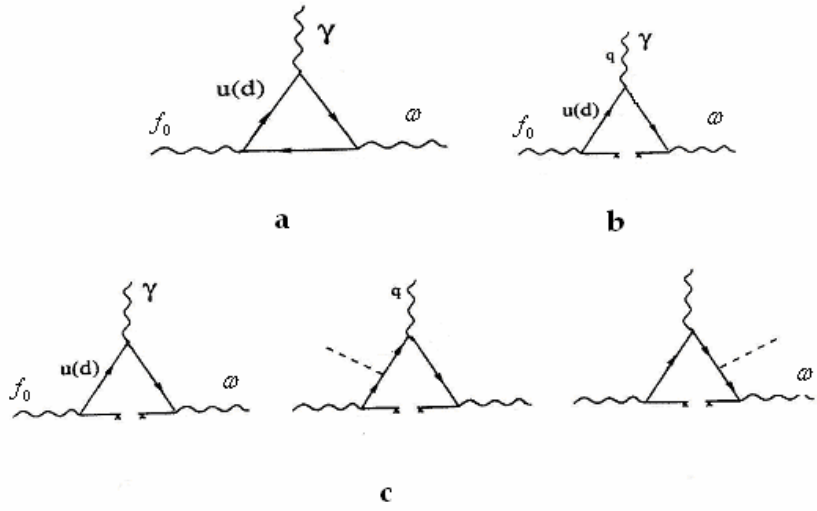

Figure 1. Feynman diagrams for the $f_0\omega\gamma$-vertex: a) bare loop diagram, b) d=3 operator corrections, and c) d=5 operator corrections. The dotted lines denote gluons.

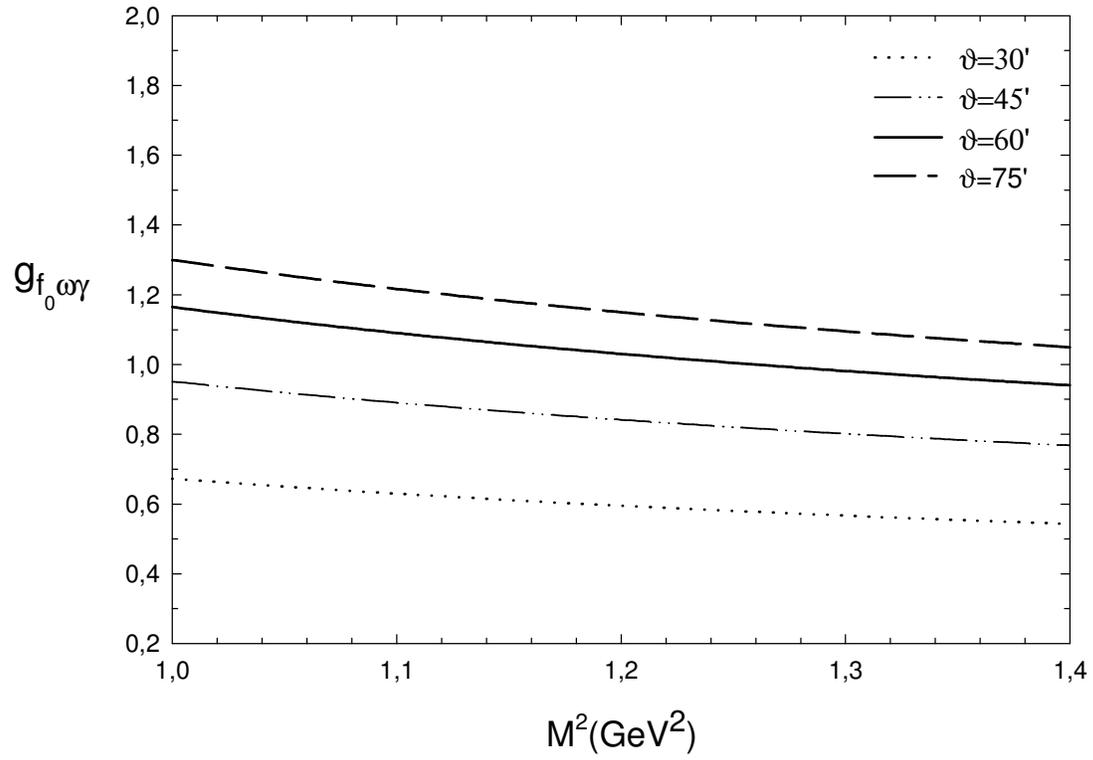

Figure 2. The variation of the coupling constant $g_{f_0\omega\gamma}$ as a function of Borel parameter $M^2$ at different $\theta$ values in the calculation of 3-point QCD sum rules.

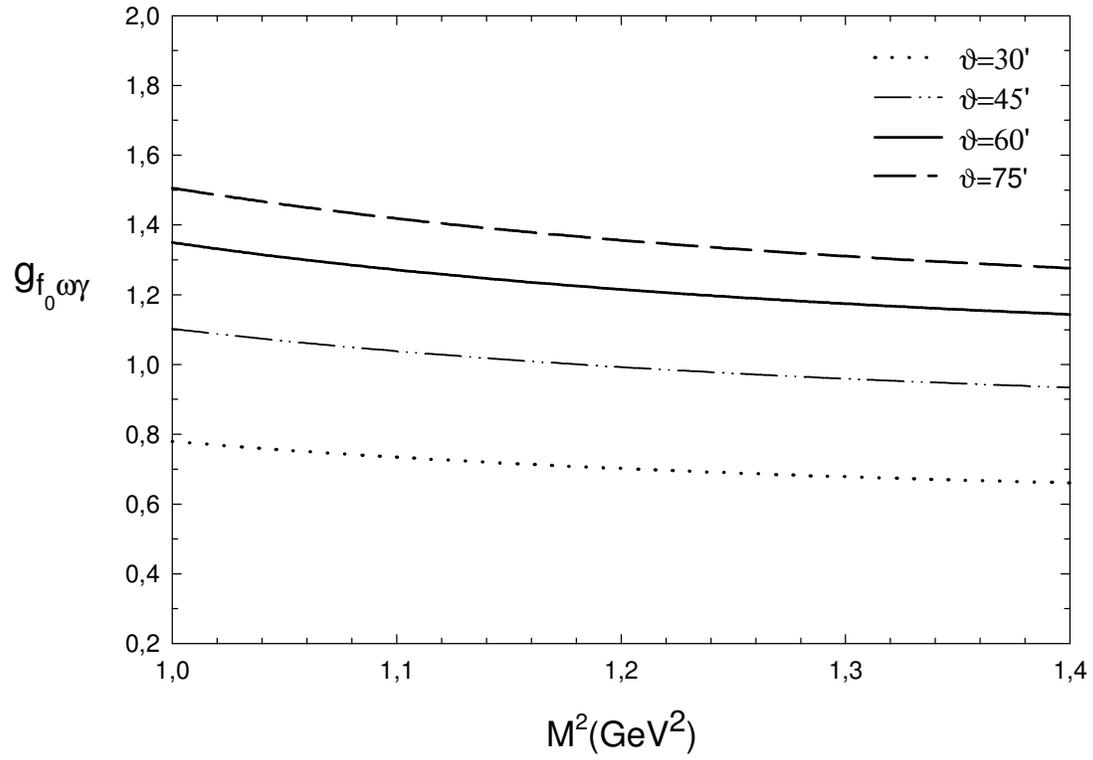

Figure 3. The dependence of the coupling constant $g_{f_0 \omega \gamma}$ on parameter $M^2$ at constant value of the continuum threshold as $s_0 = 2.0$ in the light-cone sum rules calculation.

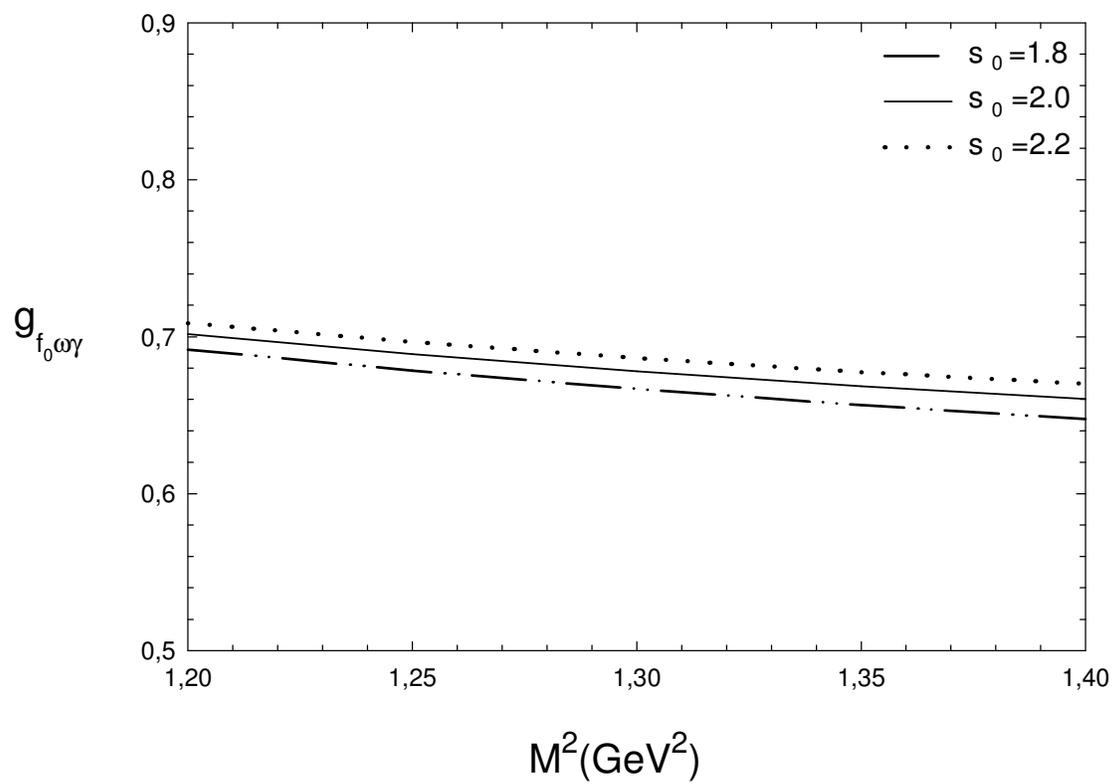

Figure 4. The dependence of the coupling constant $g_{f_0\omega\gamma}$ on parameter $M^2$ at some different values of the continuum threshold as $s_0 = 1.8,\ 2.0$ and $2.2\ \text{GeV}^2$ at $\theta = 30°$ in the light-cone sum rules calculation.

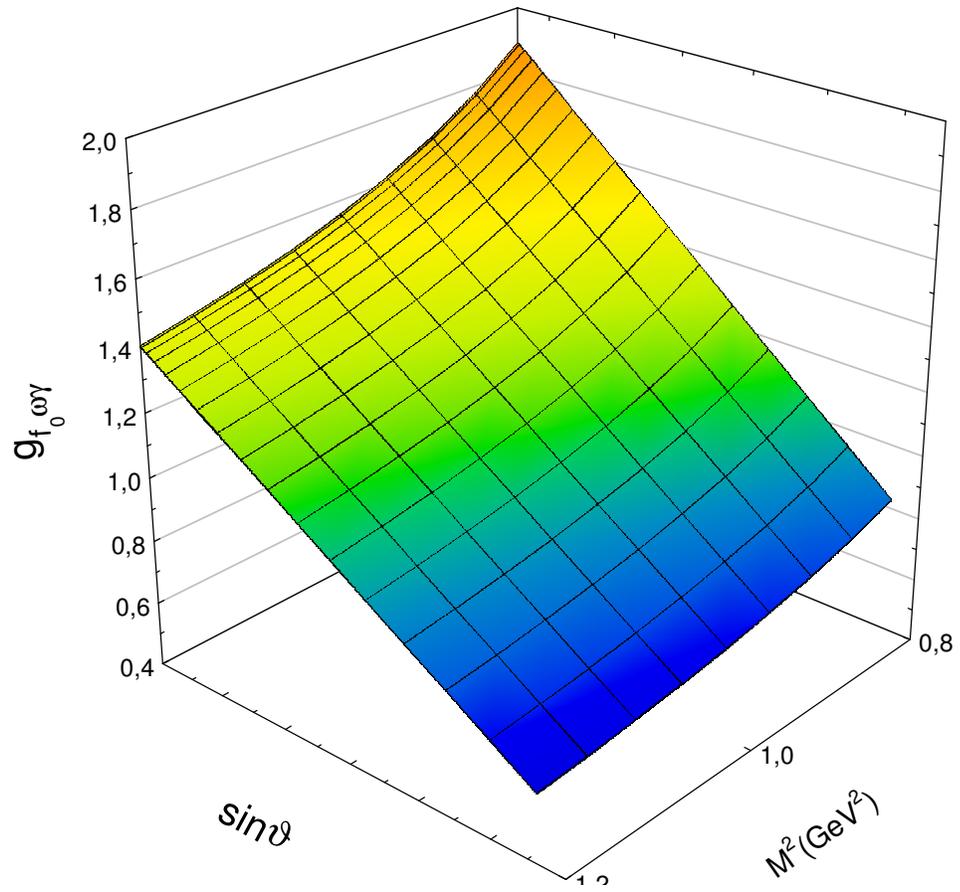

Figure 5. Coupling constant $g_{f_0 \omega \gamma}$ as function of $M^2$ and $\sin\theta$.